# Radiative Cooling with Angular Shields: Mitigating Atmospheric Radiation and Parasitic Heating

Mohamed ElKabbash[1, *]

[1.] Research Laboratory of Electronics, MIT, Cambridge, MA, 02139, USA

* Corresponding email: melkabba@mit.edu

**Abstract:**

Radiative cooling emerged as a possible sustainable solution to the energy hungry vapor compression-based cooling. However, realizing subfreezing temperatures through radiative cooling remains challenging in environments with high humidity and often requires extreme heat management, e.g., by placing the thermal emitter in ultrahigh vacuum conditions. This work theoretically investigates the introduction of angular selective thermal emission through surrounding the emitter with an angular shield. The effect of the spectral selectivity of the emitter, the humidity of the environment, and the introduction of parasitic heating on the cooling performance is studied. The optimal angle for the shield under ideal conditions is shown to be 45°. In addition, spectral selectivity of thermal emission is necessary to obtain noticeable improvement in the minimum equilibrium temperature. In humid environments, angular selectivity through engineering the thermal emissivity function of the emitter provides a better cooling performance compared to angular shields. Conversely, angular shields performance is superior when introducing parasitic heating. Using angular shields enables cooling emitters to subfreezing temperatures without vacuum and under humid levels higher than the global average.

## Introduction:

To realize net zero emission of greenhouse gases, the scientific, technological, and industrial communities must search for environmentally friendly alternatives to activities that heavily emit greenhouse gases. Air conditioning and refrigeration consumes 15% of the total electricity generated worldwide, mainly through burning fossil fuels[1,2]. This share in electricity consumption is expected to increase due to the increased welfare in high population density nations and the increased demand for cooling in data centers which consumes more energy than most countries[3]. Radiative cooling is a possible clean alternative to compression-based cooling[4-6].

Radiative cooling is a passive cooling method that uses the cold universe as a heat sink. Every object at a non-zero absolute temperature radiates over a wavelength range determined by its temperature according to the blackbody radiation spectrum. The emissivity $\epsilon(\lambda, \theta) = [0,1]$ of an object determines how closely it radiates as a blackbody, i.e., a perfect thermal emitter, with $\epsilon(\lambda, \theta) = 1$. According to Kirchhoff's law of thermal radiation, an object's emissivity $\epsilon(\lambda, \theta)$ is equal to its absorptivity $\alpha(\lambda, \theta)$ at a given wavelength and angle. To radiatively cool a thermal emitter with a radiation flux $Q_{emitter}$, the net radiated flux must be positive. The net radiated flux $Q_{net}$ is given by

$$Q_{net}(T) = Q_{emitter}(T) - Q_{atm}(T_{atm}) - Q_{sun} - Q_{parasitic} \quad (1)$$

where $Q_{atm}$, $Q_{sun}$, and $Q_{parasitic}$ are the thermal flux absorbed by the emitter from atmospheric radiation, solar radiation and parasitic (conductive and convective) heating, respectively. The equilibrium temperature of the thermal emitter is when $Q_{net} = 0$.

Radiative cooling was studied in the 1970s and 1980s starting with the landmark works by Trombe et. al.[7] Early works focused on spectral selectivity using materials with low absorption within the solar spectrum (~ 200 nm - 2500 nm) and high emissivity within the atmospheric transparency window (8 μm – 13 μm). Introducing spectrally selective thermal emitters led to high efficiency nighttime radiative cooling[7] and even daytime radiative cooling[8]. A revival in radiative cooling engineering efforts started with Raman et. al.[9], who realized sub-ambient daytime radiative cooling through using a combination of efficient spectrally selective thermal emitters and better control over solar absorption by using photonic crystals. Deep subfreezing radiative cooling was shown by Chen et. al.[10] by managing parasitic heating through placing a thermal emitter in a vacuum chamber. However, subfreezing daytime radiative cooling remains challenging in regions with relatively high humidity[11]. In addition, subfreezing cooling required turbo pumps which makes it impossible to scale and be energy efficient[10].

Because the atmospheric emissivity increases at higher angles[7], angular selective thermal emission was investigated as an additional design degree of freedom. Gentle and Smith[12] suggested that angular selective thermal emission can enhance the net radiated power from thermal emitters. Zhou et. al. [13] showed that adding an angular shield improves the cooling performance of thermal emitters by reducing the effect of thermal radiation from buildings and the higher emissivity of the atmosphere at higher angles measured from the zenith. Sandeep and coworkers [14] theoretically showed that angularly selective thermal emitters improve the performance of thermal emitters particularly in conditions where atmospheric transmission drops due to humidity.

Although selective thermal emission reduces the absorbed radiation from the atmosphere, it also reduces the radiated power from the thermal emitter. This means that parasitic heating become more detrimental as selectivity increases. It was argued that adding an angular shield can reduce radiation from the atmosphere without reducing the emitter's radiation[14]. This means that the effect of parasitic heating can be mitigated by using angular shields. However, this claim was not studied rigorously and did not consider the fact that the radiation from the atmosphere within the viewed angular range will increase due to reflection off the angular shield and to the sample.

This work studies the cooling performance of thermal emitters with angular selectivity imposed through angular shields. We show that a shield with a $45°$ angle provides the best cooling performance under ideal conditions. The cooling performance of adding an angular shield is worse than that of engineered thermal emissivity under ideal conditions and when operating under humid and less transmissive atmosphere. However, the cooling performance using shields is superior against parasitic heating. We show that subfreezing cooling is possible through simple control over parasitic heating without vacuum when the average atmospheric transmittance within the transparency window at normal incidence > 70%. Under high vacuum, obtaining subfreezing temperatures with angular shields is possible for an average transmittance > 40%. Because of its design simplicity and excellent performance under realistic conditions, we believe that angular shields can lead to the proliferation of radiative cooling devices.

**Theoretical Background:**

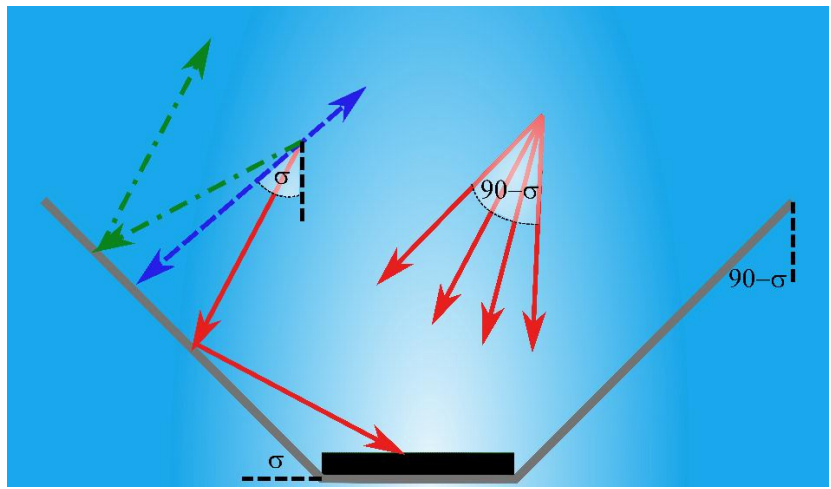

**Figure 1: Geometry of the problem:** We consider an omnidirectional thermal emitter that does not absorb within the solar spectrum and can either thermally radiate as a blackbody or selectively, i.e., within the atmospheric transparency window. The emitter is placed inside an angular shield forming an angle $\sigma$ with the horizon. The emitter absorbs thermal radiation from the atmosphere through two channels. The first is direct radiation from the atmosphere which occurs at all angles $0 < \theta < 90 - \sigma$. The second is due to radiation reflected off the shields and collected by the sample. Radiation at angles $> \sigma$ is reflected to the atmosphere by the angular shield.

**Figure 1** provides a schematic of the system under investigation. We consider a thermal emitter (black) surrounded by a conic shield acting as a perfect electric conductor with $\epsilon_{Shield}(\lambda, \theta) = 0$. The shield's angle with respect to the flat sample is $\sigma$. The emitter absorption/emission is assumed to be angle independent at all wavelengths. We assume that it is a selective absorber, i.e., it does not absorb solar radiation and its $\alpha_{Emitter}(\lambda = 200nm - 2500nm) = 0$. Consequently, $Q_{sun} = 0$. The thermal emissivity of the emitter $\epsilon_{emitter}(\lambda, \theta) = 1$ for all wavelengths within the atmospheric window (Spectrally Selective) or for all wavelengths $\lambda > 2500nm$ (Blackbody). The heat flux radiated from the emitter $Q_{emitter}$ is given by

$$Q_{emitter}(T) = \int d\Omega \cos\theta \int_0^\infty d\lambda\, I_{BB}(T,\lambda)\, \epsilon_{emitter}(\lambda, \theta) \quad (2)$$

where $\int d\Omega = 2\pi \int_0^{\pi/2} d\theta \sin\theta$ is an integral over the hemispherical solid angle and $I_{BB}(T,\lambda) = \frac{2hc^2}{\lambda^5} \frac{1}{e^{hc/\lambda k_B T} - 1}$ is the intensity of the blackbody radiation at a given temperature $T$ of the thermal emitter, $h$ is plank's constant, $c$ is the speed of light, and $k_B$ is the Boltzmann constant.

The emitter absorbs the thermally radiated heat flux from the atmosphere $Q_{atm}$ through two channels. The first is direct radiation from the atmosphere which occurs at all angles $0 < \theta < 90 - \sigma$. The second is due to radiation reflected off the shields and collected by the sample. For all angles $\theta \geq \sigma$ the radiation is reflected. Accordingly, $Q_{atm}$ is given by

$$Q_{atm}(T_{atm}) = \int_0^{90-\alpha} d\Omega \cos\theta \int_0^\infty d\lambda\, I_{BB}(T_{atm}, \lambda)\, \epsilon_{atm}(\lambda, \theta)\, \epsilon_{emitter}(\lambda, \theta) +$$
$$\int_0^\alpha d\Omega \cos\theta \int_0^\infty d\lambda\, I_{BB}(T_{atm}, \lambda)\, \epsilon_{atm}(\lambda, \theta)\, \epsilon_{emitter}(\lambda, \theta) \quad (3)$$

$I_{BB}(T_{atm}, \lambda)$ is the intensity of the blackbody radiation of the atmosphere at ambient temperature $T_{atm}$, here taken as $293\,°K$. $\epsilon_{atm}(\lambda, \theta)$ is the spectrally and angularly dependent thermal emissivity of the atmosphere. For most of the results below, $\epsilon_{atm}(\lambda, \theta) = 1 - \tau_{atm}^{1/Cos(\theta)}$, where $\theta$ is the polar angle, and $\tau_{Atm}$ is the atmospheric transmittance. The atmospheric emissivity is assumed to have no dependence on the azimuthal angle. Moreover, $\tau_{atm} = 0.8$ is assumed within the IR atmospheric transmission window ($8\,\mu m - 13\,\mu m$) and zero everywhere else for simplicity unless otherwise stated.

The heat flux absorbed by the sample due to convection and conduction, $Q_{parasitic}$, is given by:

$$Q_{parasitic} = h\,(T_{atm} - T) \quad (4)$$

where $h$ is an effective heat transfer coefficient. Under ambient conditions, $h = 8\,W.m^{-2}.K^{-1}$. Under reasonable heat management to minimize convection and conduction, $h = 2\,W.m^{-2}.K^{-1}$ is attainable [10]. For emitters in high vacuum chambers, $h = 0.2\,W.m^{-2}.K^{-1}$ was experimentally demonstrated[10].

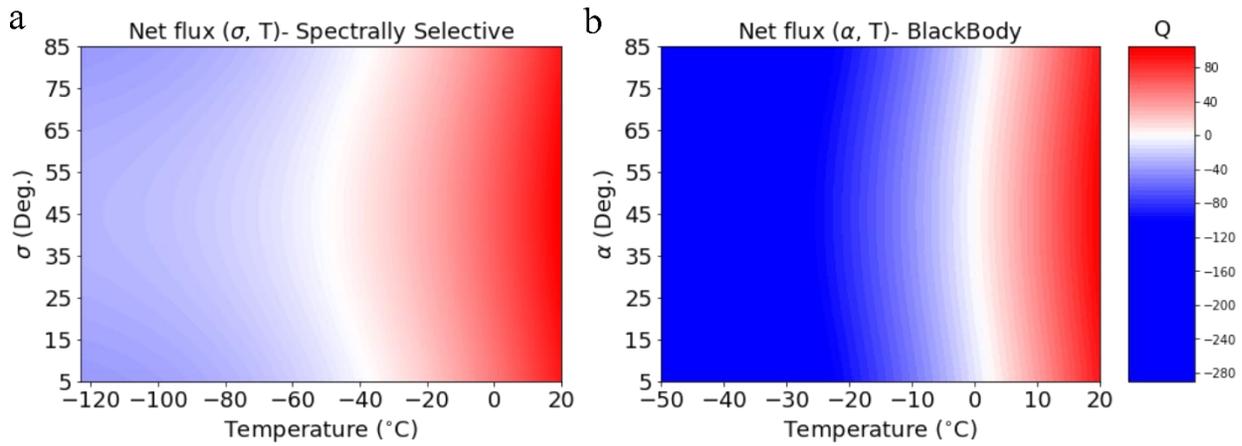

**Figure 2:** Effect of the shield's angle $\sigma$ on the steady state temperature of **(a)** a spectrally selective thermal emitter, and **(b)** a blackbody emitter. Angular selectivity plays little role in minimizing the blackbody emitter temperature. The equilibrium temperature of the emitter takes place at $Q_{net} = 0$ (white contour).

**Effect of thermal emission selectivity:**

To consider the effect of thermal spectral selectivity, we compare $Q_{net}$ as a function of the shield's angle $\sigma$ for a spectrally selective thermal emitter, i.e., $\epsilon_{emitter}(\lambda, \theta) = 1$ for $\lambda = 8\mu m - 13\mu m$ and zero everywhere, compared to a blackbody emitter with $\epsilon_{emitter}(\lambda, \theta) = 1$ at all $\lambda > 2500\ nm$. Here, we ignore the effect of parasitic heating, i.e., $Q_{parasitic} = 0$. In both cases, the lowest equilibrium temperature is reached for $\sigma = 45°$. In addition, spectral selectivity is necessary to enhance the radiative cooling performance of thermal emitters. This is because the blackbody radiation spectrum of the atmosphere overwhelms the contribution of the narrow atmospheric transparency window on the net radiated flux. The minimum temperature attainable under these conditions with a spectrally selective emitter is $-53°C$. Note that angular selective thermal emission through engineered $\epsilon_{emitter}(\lambda, \theta)$ provides a better performance under the same conditions with a steady state temperature of $-64°C$ [14].

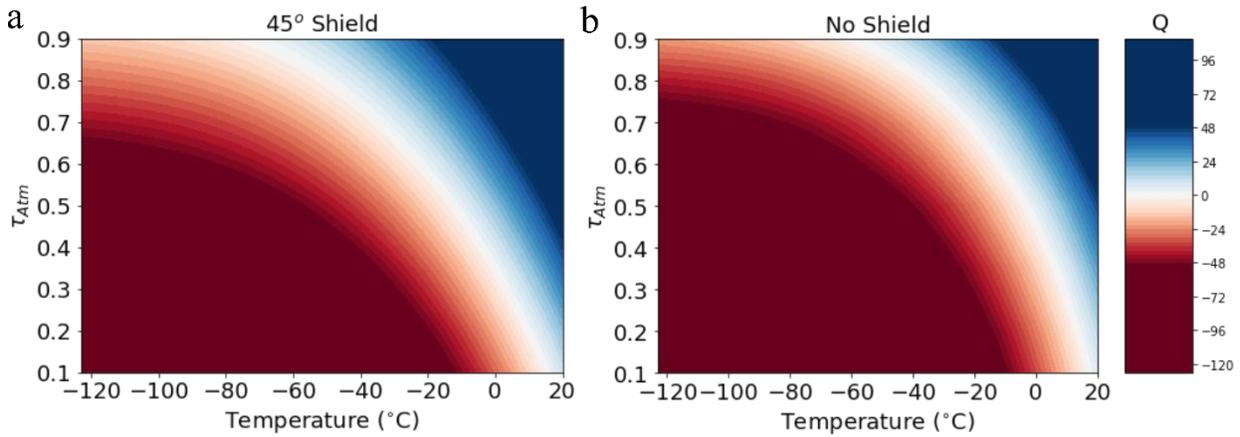

**Figure 3:** Effect of atmospheric transmission $\tau_{atm}$ on the cooling performance of a thermal emitter **(a)** with and **(b)** without an angular shield. The shield angle is $\sigma = 45°$. The equilibrium temperature of the emitter takes place at $Q_{net} = 0$ (white contour).

### Effect of atmospheric water vapor:

Water content in the atmosphere decreases its transparency significantly. Angular selectivity provides an additional degree of freedom to minimize the detrimental effects of water vapor on radiative cooling performance. In low humidity regions, e.g., polar areas, winter hemisphere, Tibetan plateau and the Middle East, the precipitable water is below 20 mm, [15] which corresponds to $\tau_{atm} > 0.8$ [6]. On the other hand, precipitable water is abundant over the tropics, Indian Ocean, Gulf of Mexico, southeastern USA and coastal China which explains why Tso et al.[11] failed to realize daytime radiative cooling using the same design introduced by Raman et al.[2] when conducting a field test in Hong Kong.

**Figure 3** shows $Q_{net}$ as a function of the atmospheric average transmittance within the atmospheric window for a spectrally selective thermal emitter surrounded by a shield at $\sigma = 45°$ (**Fig. 3a**) compared to the case without a shield (**Fig. 3b**). We also assume that $Q_{parasitic} = 0$. The cooling performance improves significantly with the shield and subfreezing temperatures are possible for average atmospheric transmission $> 0.35$, which approximately corresponds to a precipitable water of 60 mm. Note that engineered angular emission provides a better performance than angular shields for the same conditions[14].

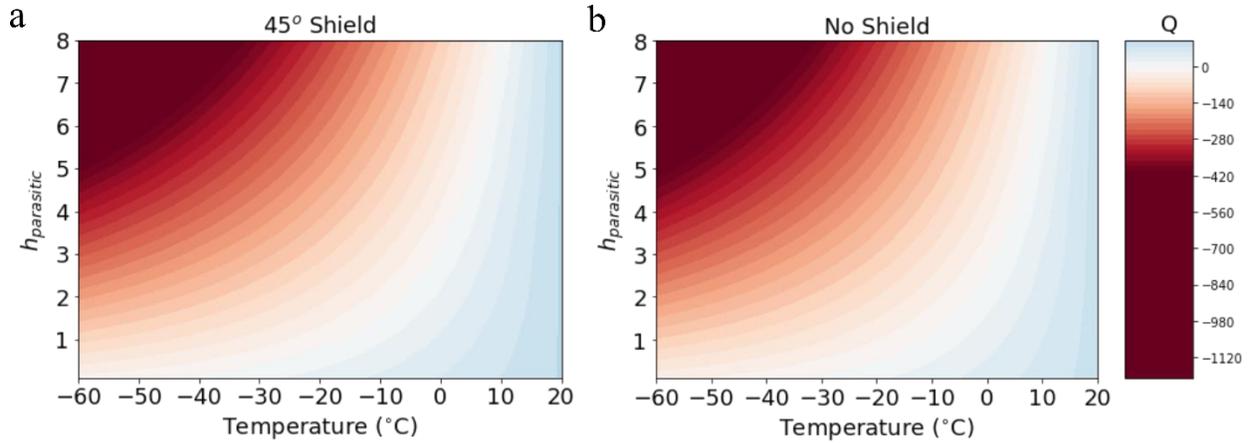

**Figure 4:** Effect of parasitic heating on the cooling performance of a thermal emitter **(a)** with and **(b)** without an angular shield. The shield angle is $\sigma = 45°$. At all parasitic heating coefficient $h$, the angular shield provides a better performance than without a shield.

## Effect of Parasitic Heating:

The advantage of angular shields over engineered angular thermal emission appears when considering the effect of parasitic heating. This is because angular shields do not restrict the angular emission of the emitter itself which can radiate at all angles efficiently which reduces the detrimental effects of parasitic heating. **Figure 4** compares $Q_{net}$ as a function of the parasitic heating coefficient $h$ for a spectrally selective thermal emitter surrounded by a shield at $\sigma = 45°$ (**Fig. 4a**) compared to the case without a shield (**Fig. 4b**). We assume that $\tau_{atm} = 0.8$. Subfreezing temperatures are possible for $h < 3.3 \ W.m^{-2}.K^{-1}$ and $h < 2.5 \ W.m^{-2}.K^{-1}$, for the shielded emitter and unshielded emitter, respectively. Angular shields, thus, improve the performance even when parasitic heating is involved. This improvement persists even at ambient conditions where $h \sim 8\ W.m^{-2}.K^{-1}$. Angular shields also provide better cooling performance compared to engineered thermal emitters particularly for high parasitic heating coefficients[14].

**Realistic design:**

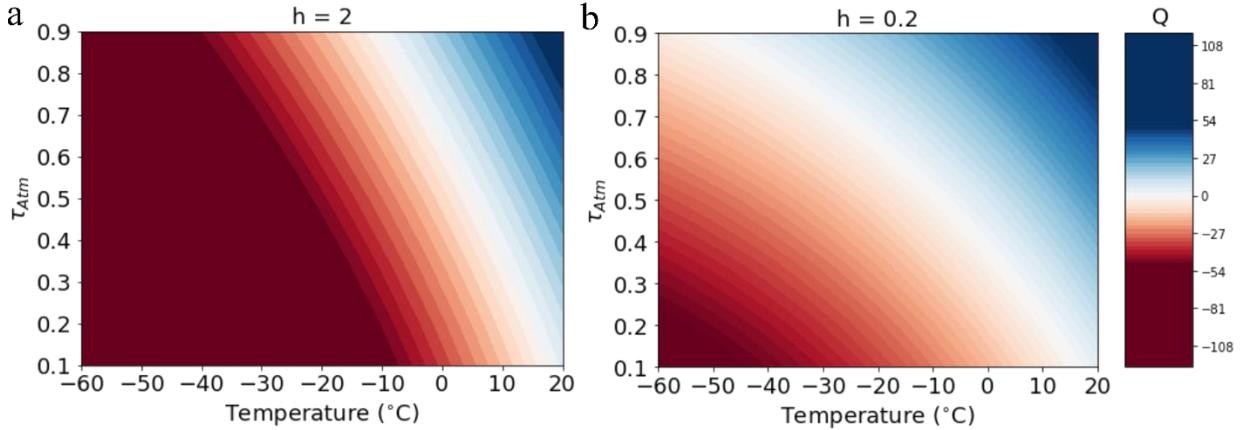

**Figure 5:** Performance of an emitter in an angular shield with $\sigma = 45°$ as a function of the atmospheric transmittance for a parasitic heating coefficient of **(a)** $h = 2\ W.m^{-2}.K^{-1}$ and **(b)** $h = 0.2\ W.m^{-2}.K^{-1}$. Subfreezig temperatures are obtained for $\tau_{atm} > 0.65$ for **(a)** and $\tau_{atm} > 0.35$ for **(b)**.

We consider the performance of thermal emitters protected by an angular shield with $\sigma = 45°$ under conditions experimentally attainable. With some heat management and without a turbo pump, a heat transfer of $h = 2\ W.m^{-2}.K^{-1}$ is certainly possible[8,14]. **Figure 5a** shows that under these conditions, subfreezing cooling can take place for $\tau_{atm} > 0.65$ which corresponds to a precipitable water of ~30 mm, i.e., higher than the global average of 21.6 mm[15].

A heat transfer coefficient of $h = 0.2\ W.m^{-2}.K^{-1}$ is possible but requires a turbo pump[10]. Under these conditions, subfreezing temperatures are possible for atmospheric transmission $\tau_{atm} > 0.37$. Note that the dimensions of the shield should be such that no direct thermal radiation from the atmosphere are absorbed by the emitter at angles $> 90° - \sigma$.

**Conclusions:**

In summary, angular selectivity, whether through adding an angular shield or by engineering the angular thermal emissivity, improve the cooling performance of thermal emitters. Angular selectivity through shields enjoys two distinct features, (i) the shield reflects thermal emission from the atmosphere incident at angles that are generally different from the angles blocked by the shield and (ii) the thermal emitter can radiate at all angles.

Due to its first feature, engineered thermal emission provides a better cooling performance particularly under opaque atmospheric conditions which takes place in humid environments. Due to the second feature, however, angular shields provide a significantly better cooling performance when we consider parasitic heating. In addition, angular shields are simple to design and implement compared to engineered angular thermal emission which requires the deposition of multilayers thin films[14,16,17].

With properly designed angular shields, subfreezing temperatures under realistic conditions are possible to achieve with and without vacuum turbo pumps which is extremely promising. By considering the angular degree of freedom in designing thermal emitters, radiative cooling could find widespread real-world applications and contribute significantly to humanity's struggle towards a sustainable future.

## References:

1   Bellemo, L. e. a. Modelling and analysis of a desiccant cooling system using the regenerative indirect evaporative cooling process. *ECOS 2013, Proceedings of the 26th International Conference on Efficiency, Cost, Optimization, Simulation and Environmental Impact of Energy Systems: Guilin, China*, 109-122 (2013).
2   Chu, S. & Majumdar, A. Opportunities and challenges for a sustainable energy future. *Nature* **488**, 294, doi:10.1038/nature11475 (2012).
3   Metzger, I., Dean, J. & Kurnik, C. Psychrometric bin analysis for alternative cooling strategies in data centers. *ASHRAE Transactions* **117**, 254 (2011).
4   Li, W. & Fan, S. Radiative cooling: harvesting the coldness of the universe. *Optics and Photonics News* **30**, 32-39 (2019).
5   Yin, X., Yang, R., Tan, G. & Fan, S. Terrestrial radiative cooling: Using the cold universe as a renewable and sustainable energy source. *Science* **370**, 786-791 (2020).
6   Zhao, D. *et al.* Radiative sky cooling: Fundamental principles, materials, and applications. *Applied Physics Reviews* **6**, 021306 (2019).
7   Granqvist, C. & Hjortsberg, A. Radiative cooling to low temperatures: General considerations and application to selectively emitting SiO films. *Journal of Applied Physics* **52**, 4205-4220 (1981).
8   Harrison, A. & Walton, M. Radiative cooling of TiO2 white paint. *Solar Energy* **20**, 185-188 (1978).
9   Raman, A. P., Anoma, M. A., Zhu, L., Rephaeli, E. & Fan, S. Passive radiative cooling below ambient air temperature under direct sunlight. *Nature* **515**, 540-544 (2014).
10  Chen, Z., Zhu, L., Raman, A. & Fan, S. Radiative cooling to deep sub-freezing temperatures through a 24-h day–night cycle. *Nature communications* **7**, 1-5 (2016).

11  Tso, C. Y., Chan, K. C. & Chao, C. Y. A field investigation of passive radiative cooling under Hong Kong's climate. *Renewable energy* **106**, 52-61 (2017).
12  Gentle, A. & Smith, G. in *Nanostructured Thin Films II.* 96-103 (SPIE).
13  Zhou, L. *et al.* A polydimethylsiloxane-coated metal structure for all-day radiative cooling. *Nature Sustainability* **2**, 718-724 (2019).
14  Chamoli, S. K., Li, W., Guo, C. & ElKabbash, M. Angularly selective thermal emitters for deep subfreezing daytime radiative cooling. *Nanophotonics*, doi:doi:10.1515/nanoph-2022-0032 (2022).
15  Liang, J. *Chemical modeling for air resources: fundamentals, applications, and corroborative analysis.* (Academic Press, 2013).
16  Shen, Y. *et al.* Optical broadband angular selectivity. *Science* **343**, 1499-1501 (2014).
17  Qu, Y. *et al.* Polarization-Independent Optical Broadband Angular Selectivity. *ACS Photonics* **5**, 4125-4131, doi:10.1021/acsphotonics.8b00862 (2018).